# Interfacial and bulk spin-Hall contributions to field-like spin-orbit torque generated by Iridium

Sutapa Dutta[1,2], Arnab Bose*[1], A. A. Tulapurkar[2], R. A. Buhrman[1] and D. C. Ralph[3,4]

1 School of Applied and Engineering Physics, Cornell University, New York, US 14853
2 Dept. of Electrical Engineering, Indian Institute of Technology Bombay, Mumbai, India 400076
3 Laboratory of Atomic and Solid State Physics, Cornell University, New York 14853, USA.
4 Kavli Institute at Cornell for Nanoscale Science, Ithaca, New York 14853

*ab2729@cornell.edu

**Abstract**

We present measurements of spin-orbit torques (SOTs) generated by Ir as a function of film thickness in sputtered Ir/CoFeB and Ir/Co samples. We find that Ir provides a damping-like component of spin-orbit torque with a maximum spin torque conductivity $\sigma_{DL}^{eff}=(1.4 \pm 0.1) \times 10^5 \frac{\hbar}{2e}\Omega^{-1}m^{-1}$ and a maximum spin torque efficiency of $\xi_{DL} = 0.042 \pm 0.005$, which is sufficient to drive switching in an 0.8 nm film of CoFeB with perpendicular magnetic anisotropy. We also observe a surprisingly large field-like spin-orbit torque (FLT). Measurements as a function of Ir thickness indicate a substantial contribution to the FLT from an interface mechanism, so that in the ultrathin limit there is a non-zero FLT with a maximum torque conductivity $\sigma_{FL}^{eff} = -(5.0 \pm 0.5) \times 10^4 \frac{\hbar}{2e}\Omega^{-1}m^{-1}$. When the Ir film thickness becomes comparable to or greater than it's spin diffusion length, 1.6 ± 0.3 nm, there is also a smaller bulk contribution to the field-like torque.

**Introduction**

Spin-orbit interactions (SOIs) play a central role in the development of next-generation spintronic devices for non-volatile magnetic random-access memory (MRAM) applications [1,2]. Heavy metals with large SOI such as Pt [3,4], Ta [5–7], and W [8–10] are efficient generators of spin current to produce spin-orbit torques (SOTs) on an adjacent magnetic layer. Iridium (Ir) is another heavy metal with a large SOI [11], which has been used to realize skyrmions and chiral domain wall structures as it can provide a strong interfacial Dzyaloshinskii-Moriya interaction (DMI) [12]. Recently it has been reported that Ir is also capable of converting charge current into spin current, with a damping-like torque efficiency $\xi_{DL}$ = 0.005-0.01 (where $\xi_{DL} \equiv \tau_{DL}/J_C = T_{int}\theta_{SH}$, with $\tau_{DL}$ the spin-orbit torque per unit area, $J_C$ the applied current density, $T_{int}$ an interface spin transparency factor, and $\theta_{SH}$ the spin Hall ratio within the heavy metal), as measured by current-induced domain wall motion [13], hysteresis loop shifts [14], and second-harmonic Hall measurements [15], and $\theta_{SH} \approx 0.02$ as measured by spin pumping [16,17]. Magnetic switching driven by spin-orbit torque from Ir has also been observed [15,18]. Here, we use spin-torque ferromagnetic resonance (ST-FMR) and 2nd-harmonic Hall (SHH) measurements to confirm that Ir exhibits a sizable charge-to-spin-current conversion efficiency, with a maximum damping-like spin torque conductivity, $\sigma_{DL}^{eff}= (1.4 \pm 0.1) \times 10^5 \frac{\hbar}{2e}\Omega^{-1}m^{-1}$ (where $\sigma_{DL}^{eff} \equiv \frac{\hbar}{2e}\xi_{DL}/\rho_{Ir}$, with $\rho_{Ir}$ the resistivity of the Ir) and a maximum $\xi_{DL}$ of 0.042 ± 0.005, slightly higher than previous reports. We also observe a substantial field-like torque (FLT). By varying the thickness of Ir from 1.2-5 nm we find a dominant interface contribution to the FLT, in addition to a smaller conventional bulk contribution that grows with film thickness on the scale of the spin diffusion length.

Spin-current generation and spin-orbit torques in heavy metal(HM)/ferromagnet(FM) structures can in principle arise from either a bulk spin Hall effect (SHE) in the HM or interfacial SOI present at the interface [1,2]. Either type of effect can produce both a damping-like torque (DLT) ($\boldsymbol{\tau}_{DL} \propto \boldsymbol{m} \times \boldsymbol{\sigma_y} \times \boldsymbol{m}$) and a field-like torque (FLT) ($\boldsymbol{\tau}_{FL} \propto \boldsymbol{m} \times \boldsymbol{\sigma_y}$) (where $\boldsymbol{\sigma_y}$ is the orientation of a current-induced spin, in-plane and perpendicular to the current). First-principles calculations generally suggest that the bulk SHE should produce a DLT greater than the FLT since the real part of the interfacial spin mixing conductance ($G_r$) is larger than the imaginary part ($G_i$) [19,20]. Experimentally, it is possible to separate out the interface-driven and the bulk SHE driven SOTs by varying the thickness of the HM [21,22]. We model the thickness dependence of the spin-orbit torque conductivity as approximately (this form assumes that spin back flow is not strongly thickness-dependent) [23,24]:

$$\sigma_{SOT} = \sigma_{SOT}^{interface} + \sigma_{SOT}^{bulk}\left(1 - \text{sech}\frac{t_{HM}}{\lambda_{HM}}\right). \tag{1}$$

We have measured four different sets of samples. Set 1: Ti(1)/Ir(3)/Co(2-6)/Ta(1.2), Set 2: Ti(1)/Ir(1-5)/CoFeB(2.3)/Ta(1.2), Set 3: Ti(1)/Ir(1-5)/Co(2.3)/Ta(1.2) and Set 4: Ti(1)/Ir(1-5)/CoFeB(2.3)/Ir(1.2)/Ta(1.2). (The numbers in parentheses are thicknesses in nanometers and the stoichiometry of the CoFeB layers is $Co_{20}Fe_{60}B_{20}$.) The heterostructures are deposited by magnetron sputtering using an Ar pressure of 2 mTorr, in a system with a base pressure less than $2 \times 10^{-8}$ Torr. The 1 nm Ti layer acts as a seed layer to give smooth films, and the 1.2 nm Ta layer provides capping to prevent oxidation of the magnetic layer. Optical lithography and argon ion milling are used to pattern $30 \times 20$ μm² wires with microwave-compatible electrodes for ST-FMR measurements and Hall bars ($20 \times 6$ μm²) for the SHH and switching experiments. Devices of Set 1 are used for the ST-FMR measurement [25] to quantify the spin torque ratios for the DLT ($\xi_{DL}$) and FLT ($\xi_{FL}$) for 3 nm thick Ir. We use Sets 2 and 3 to perform SHH measurements [26–28] of the SOTs over a wide range of Ir thickness for a fixed thickness of FM. Set 4 provides experimental controls, to verify the distinct contributions of interfacial and bulk SOTs. All measurements are performed at room temperature.

**ST-FMR measurements for 3 nm Ir films**

For the ST-FMR measurements [23,25], a radio frequency (rf) current is applied which drives the magnet into resonance. Due to mixing of the oscillating current and the time-varying anisotropic magnetoresistance, a dc voltage is generated which is measured by a lock-in amplifier as a function of external in-plane magnetic field sweep at an angle $\phi$ relative to the current direction (typically $\phi = 45°$). ST-FMR signals obtained for different thicknesses of Co in Ir(3)/Co samples are shown Fig. 1(a-c). These can be well-fit by the sum of a symmetric Lorentzian ($V_S$) (red curve in Fig. 1(a-c)) and anti-symmetric Lorentzian ($V_A$) (blue curve in Fig. 1(a-c)) plus a field-independent background, with $V_S = S\left(\frac{\Delta^2}{(H-H_0)^2+\Delta^2}\right)$, $V_A = A\left(\frac{(H-H_0)\Delta}{(H-H_0)^2+\Delta^2}\right)$. The amplitudes $S$ and $A$ are related to the in-plane and out-of-plane current induced torques, $\Delta$ is the linewidth, and $H_0$ is the resonant field.

We extract the damping-like and field-like spin torque efficiencies from the ST-FMR measurements using the method of ref. [25]. For each sample with a different Co thickness, we first define an intermediate parameter $\xi_{\text{FMR}} \equiv \frac{S}{A}\left(\frac{e}{\hbar}\right)\mu_0 M_s t_{\text{FM}} t_{Ir}\sqrt{1 + (H_\perp/H_0)}$, where $\mu_0$ is the permeability, $M_S$ is the saturation magnetization as measured by vibrating sample magnetometry, $t_{FM}$ is the Co thickness, $t_{Ir}$ is the Ir thickness, and $H_\perp$ is the out of plane demagnetization field determined from the resonant field. (Values of $M_S$ and $H_\perp$ are plotted in Supplementary Material [29].) Assuming that the spin-torque efficiencies do not depend on the thickness of the Co layer in the range we study, the damping-like and field-like spin torque efficiencies can then be determined via a linear fit of $1/\xi_{FMR}$ to $1/t_{FM}$ [25] (Fig. 1(d))

$$\frac{1}{\xi_{FMR}} = \frac{1}{\xi_{DLT}}\left(1 + \frac{\hbar}{e}\frac{\xi_{FLT}}{\mu_0 M_s t_{FM} t_{Ir}}\right) \quad (2)$$

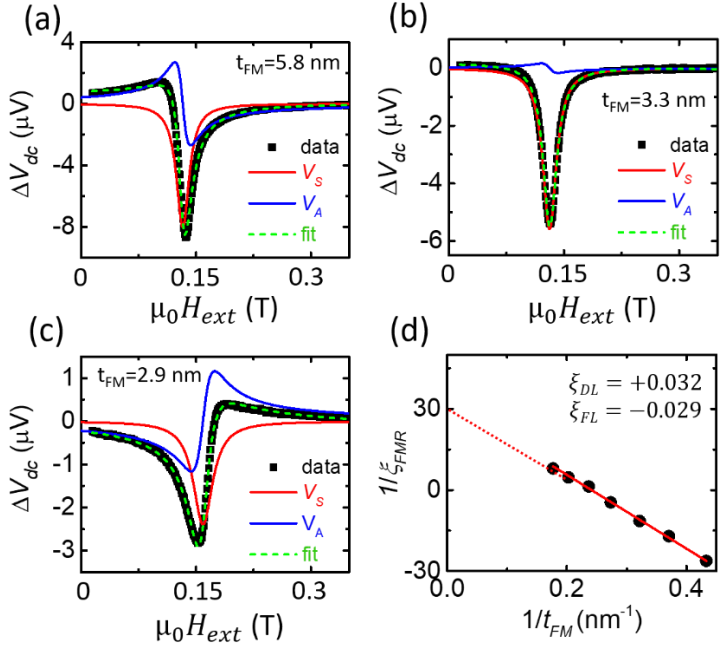

Fig. 1. (a-c) ST-FMR voltage signals for Ir (3 nm)/Co ($t_{FM}$) samples measured at a microwave frequency $f$ = 12 GHz, for samples with $t_{FM}$ = 5.8 nm, 3.3 nm, and 2.9 nm. A constant background offset voltage is subtracted in each panel. Red lines show fits to the symmetric Lorentzian component, blue lines are fits to the antisymmetric component, and green dotted lines the total fit. (d) Fit to Eq. (2) used to determine the spin-torque efficiencies.

Based on this analysis, we find $\xi_{DL}$ = 0.032±0.004 and $\xi_{FL}$ = -0.029±0.003 for the Ir(3 nm)/Co samples. The positive sign of the DLT denotes the same sign as in Pt, while the negative sign of the FLT indicates that it is opposite to the average torque produced by the in-plane Oersted field. We find it surprising that Ir generates such a large FLT, comparable to W [30,31] even though the DLT in Ir is much smaller than Pt [24,25] and W [8,30]. The large size of the FLT is evident already from the raw data in Fig. 1, in that the sign of the antisymmetric ST-FMR component is opposite in the 2.9 nm Co sample compared to the 3.3 nm and 5.8 nm samples, indicating that in the 2.9 nm sample the spin-orbit FLT is stronger than the average torque due to the Oersted field (the torque from the Oersted field is proportional to the FM thickness while the spin-orbit torque is independent of FM thickness). The same results can be expressed in terms of effective spin torque conductivities of the DLT $\sigma_{DL}^{eff} = \frac{\hbar}{2e}\xi_{DL}/\rho_{Ir}$ and FLT $\sigma_{FL}^{eff} = \frac{\hbar}{2e}\xi_{FL}/\rho_{Ir}$, where $\rho_{Ir}$= 49× $10^{-8}$ Ω-m is the resistivity of the 3 nm Ir film in this bilayer. We find $\sigma_{DL}^{eff}$ = (6.5 ± 0.8) × $10^4$ $\frac{\hbar}{2e}\Omega^{-1}m^{-1}$ and $\sigma_{FL}^{eff}$ = (-5.9 ± 0.8) × $10^4$ $\frac{\hbar}{2e}\Omega^{-1}m^{-1}$ for the 3 nm Ir samples.

**Second-harmonic Hall measurements to determine the dependence of the torque efficiencies on Ir thickness**

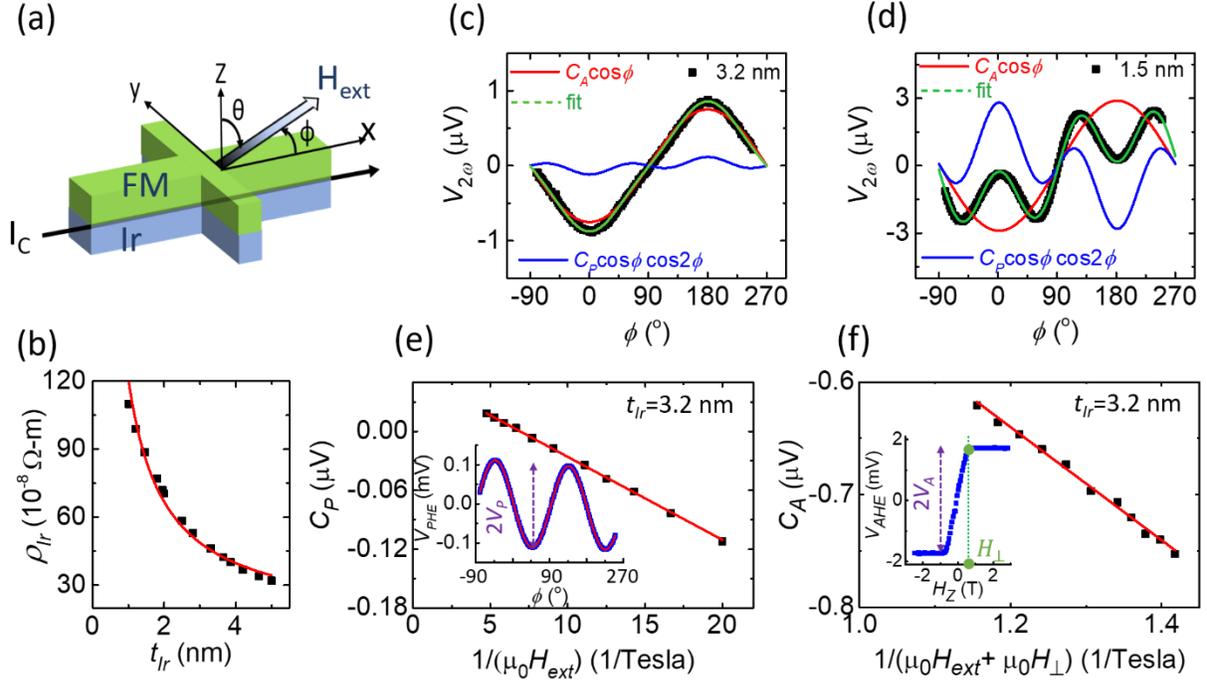

Fig. 2. (a) Sample geometry for the second-harmonic Hall measurements. (b) Ir resistivity ($\rho_{Ir}$) as a function of $t_{Ir}$. (c-d) Second-harmonic Hall voltage $V_{2\omega}$ as a function of the in-plane field angle for $\mu_0 H_{ext} = 0.05$ Tesla, for (c) $t_{Ir} = 3.2$ nm and (d) $t_{Ir} = 1.5$ nm. (e) Variation of $C_P$ as a function of $1/(\mu_0 H_{ext})$. Inset of (e): Calibration of the planar Hall effect voltage ($V_{PHE}$) with a fit to $\sin 2\phi$. (f) Variation of $C_A$ as a function of $1/(\mu_0 H_{ext} + \mu_0 H_\perp)$. Inset of (f): the anomalous Hall voltage ($V_{AHE}$) as a function of Z-field sweep.

In the SHH measurements [26–28], a low-frequency (1327 Hz in our experiment) ac current is applied and the Hall voltages (both 1st and 2nd harmonic) are measured using a lock-in amplifier as a function of rotating the applied magnetic field $H_{ext}$ in the sample plane (Fig. 2(a)). Effective magnetic fields $H_{Oe}^Y$, $H_{FL}^Y$ and $H_{DL}^Z$ corresponding to the current-induced torques can be determined based on the amount of current-induced tilting of the magnetization as measured by the 2nd-harmonic Hall voltage ($V_{2\omega}$) [26–28], where $H_{Oe}^Y$ is the Oersted field and $H_{FL}^Y$, $H_{DL}^Z$ are effective fields corresponding to the FLT and DLT, respectively. $V_{2\omega}$ has the form [26–28]:

$$V_{xy}^{2\omega} = C_P \cos 2\phi \cos \phi + C_A \cos \phi \qquad (3)$$

with $C_P = -(H_{FL}^Y + H_{Oe}^Y)\dfrac{V_P}{H_{ext}}$ and $C_A = -H_{DL}^Z \dfrac{V_A}{2(H_{ext}+H_\perp)} + V_{ANE} + V_{ONE} H_{ext}$

Here $V_{ANE}$ and $V_{ONE}$ are constants describing the strength of the anomalous Nernst effect (ANE) and the ordinary Nernst effect (ONE), both of which can arise due to an out-of-plane thermal gradient. $V_P$ is the coefficient of the planar Hall effect voltage, $V_{PHE} = V_P \sin 2\phi$, where $\phi$ is the in-plane angle between the magnetization ($M$) and the current flow direction, and $V_A$ is the coefficient of the anomalous Hall voltage, $V_{AHE} = V_A \cos\theta$, where $\theta$ is the angle between $M$ and the out-of-plane axis. We determine the value of $V_P$ from the first-harmonic Hall signal by rotating $H_{ext}$ in the plane (inset of Fig. 2(e)) and we obtain $V_A$ and $H_\perp$ by sweeping $H_{ext}$ out of plane (inset of Fig. 2(f)). We compute the Oersted field from Ampere's law, $H_{Oe}^Y = \frac{1}{2}J_{Ir}t_{Ir}$ where $J_{Ir}$ is the current density within the Ir layer and $t_{Ir}$ is the Ir thickness. The current density $J_{Ir}$ is calculated in a parallel-resistor model taking into account the thickness-dependent resistivity $\rho_{Ir}$ of the Ir layers (Fig. 2(b)) as determined by four-probe measurements of the resistance of Ir($t_{Ir}$)/CoFeB(2.3) from which the resistance of the CoFeB layer ($\rho_{CoFeB} \sim 140\times10^{-8}$ Ω-m) is subtracted. We find good consistency between these measurements of $\rho_{Ir}$ from the Ti(1nm)/Ir($t_{Ir}$)/CoFeB(2.3) samples and direct measurements of Ti(1nm)/Ir($t_{Ir}$) bilayers.

Figures 2(c) and 2(d) show representative 2nd harmonic data for samples with Ir thicknesses of 3.2 nm and 1.5 nm, for $\mu_0 H_{ext} = 0.05$ Tesla. The red and blue lines represent fits to $C_A \cos\phi$ and $C_P \cos 2\phi \cos\phi$ as a function of the field angle $\phi$. $H_{FL}^Y + H_{Oe}^Y$ is determined from a fit to the magnetic-field dependence of $C_P$ (Fig. 2(e)) and then the calculated Oersted field is subtracted, while $H_{DL}^Z$ is determined from a fit to the magnetic-field dependence of $C_A$ (Fig. 2(f)). We can express the final results in terms of the effective spin-torque conductivities $\sigma_{DL}^{eff}$ and $\sigma_{FL}^{eff}$ [24,25]

$$\sigma_{DL}^{eff} = \frac{2e}{\hbar}\mu_0 M_S t_{FM} \frac{H_{DL}^Z}{E} \qquad (4)$$

$$\sigma_{FL}^{eff} = \frac{2e}{\hbar}\mu_0 M_S t_{FM} \frac{H_{FL}^Y}{E} \qquad (5)$$

where $E$ is the longitudinal applied electric field.

It is evident already from Fig. 2(c,d) that when Ir thickness is reduced from 3.2 nm to 1.5 nm the strength of $H_{FL}^Y + H_{Oe}^Y$ grows dramatically relative to $H_{DL}^Z$, as the magnitude of the $\cos\phi \cos 2\phi$ component grows relative to the $\cos\phi$ component. $H_{FL}^Y + H_{Oe}^Y$ also changes the sign when the Ir thickness is around 2.8 nm for Ir/CoFeB and 3 nm for Ir/Co (see supplementary materials). Like the sign change as a function of FM thickness in the ST-FMR measurements (Fig. 1(a-c)), this can be understood as arising from a competition between the Oersted torque that scales with the Ir layer thickness and a substantial field-like SOT with a much weaker dependence.

The results of the SHH measurements are shown as a function of Ir thicknesses in Fig. 3 for the Ir/CoFeB(2.3 nm) sample series (left column), for the Ir/Co(2.3 nm) sample series (middle column), and for Ir/CoFeB(2.3 nm)/Ir(1.2 nm) control samples (right column). The behavior of the damping-like spin-torque conductivities (Fig. 3(a-c)) is as expected for a bulk spin Hall effect – the spin-torque conductivity goes to zero as the Ir thickness goes to zero following a reasonable fit to Eq. (1) with zero interface contribution. The values of $\sigma_{DL}^{eff}$ for both the Ir(3 nm)/CoFeB(2.3 nm) and the Ir(3 nm)/Co(2.3 nm) are in good quantitative agreement with the ST-FMR results for the 3 nm Ir samples. At large Ir thickness, $\sigma_{DL}^{eff}$ saturates to the values $(1.1 \pm 0.1) \times 10^5 \frac{\hbar}{2e}\Omega^{-1}m^{-1}$ for Ir/CoFeB and $(1.4 \pm 0.1) \times 10^5 \frac{\hbar}{2e}\Omega^{-1}m^{-1}$ for Ir/Co samples. This difference may be due to a difference in spin transparency factor for the two interfaces. Since the Ir resistivity ($\rho_{Ir}$) is a strong function of the Ir thickness and the spin diffusion length $\lambda_{HM}$ is expected to depend on $\rho_{Ir}$ [24], Eq. (1) is likely not strictly accurate for a single fixed value of $\lambda_{Ir}$. With this caveat, however, simple fits to Eq. (1) yield approximate spin diffusion lengths of $1.3 \pm 0.2$ nm for the Ir/CoFeB(2.3 nm) sample series and $1.9 \pm 0.3$ nm for Ir/Co(2.3 nm). Previous spin-pumping and SMR experiments found values $\lambda_{Ir}$ ~0.5-1.2 nm [14,16,17], which might depend on the resistivity and thickness [24] of the Ir films. For the Ir($t_{Ir}$ > 1.2 nm)/CoFeB(2.3 nm)/Ir(1.2 nm) control samples, $\sigma_{DL}^{eff}$ is reduced relative to the samples with a single Ir layer and goes to zero when both Ir layers have the same thickness of 1.2 nm. We ascribe this to cancelation of the contributions to the SOTs from the top and bottom Ir layers. Hence the fit starts from $t_{Ir}$=1.2 nm.

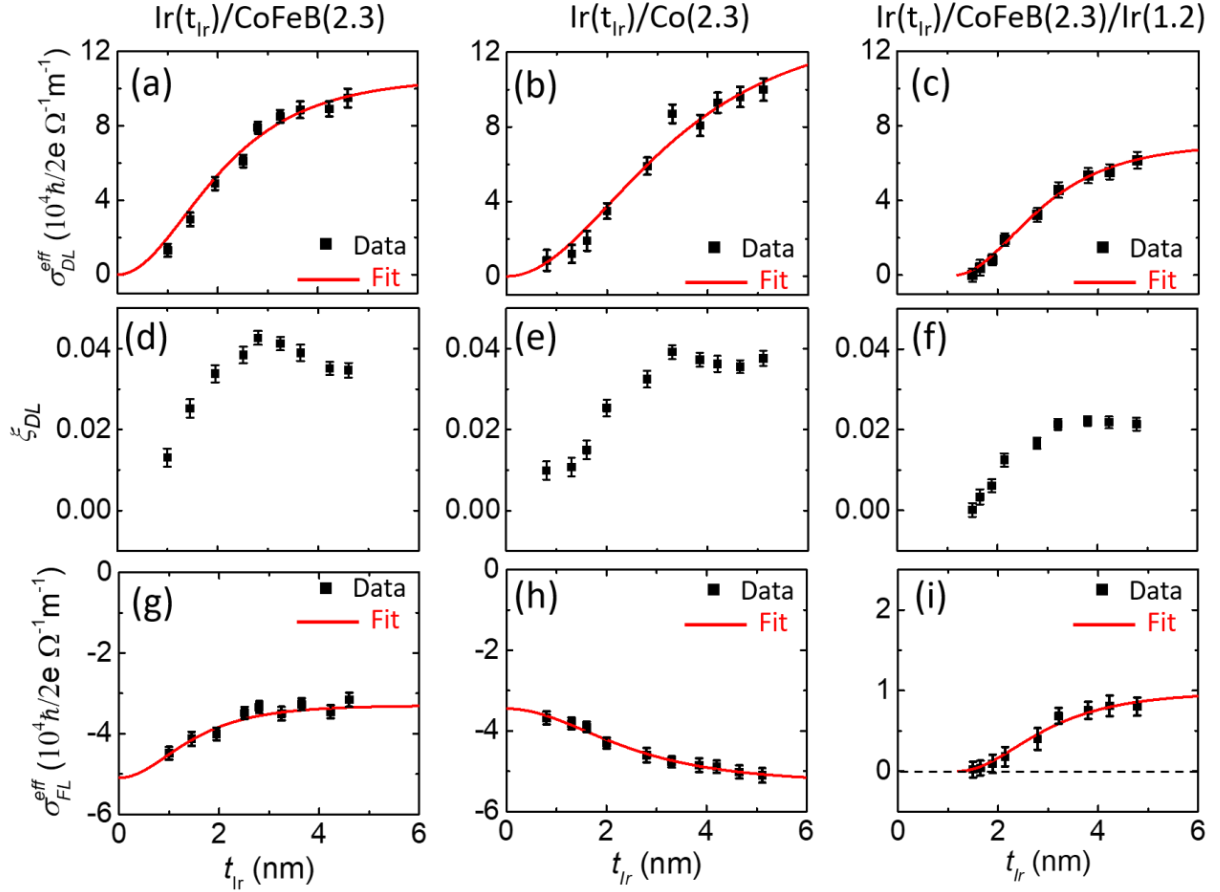

Fig. 3. Results of the second-harmonic Hall measurements of the spin-orbit torques due to Ir. (a-c) Dependence of the effective damping-like spin torque conductivity on Ir thickness for (a) the Ir/CoFeB(2.3 nm) series of samples, (b), the Ir/Co(2.3 nm) series, and (c) the Ir/CoFeB(2.3 nm)/Ir(1.2 nm) series. (d-f) Corresponding damping-like spin-torque ratio as a function of Ir thickness for the three sample series. (g-i) Dependence of the effective field-like spin torque conductivity of Ir thickness for three sample series.

Figure 3(g-h) shows the SHH results for the field-like spin torque conductivity ($\sigma_{FL}^{eff}$) as a function of Ir thickness, along with fits to Eq. (1) using the same values of $\lambda_{Ir}$ determined from the fits to $\sigma_{DL}^{eff}$. In striking contrast to the damping-like torque conductivity, $\sigma_{FL}^{eff}$ exhibits a dominant interface contribution, with values $\sigma_{FL}^{interface} = -(5.0 \pm 0.5) \times 10^4 \frac{\hbar}{2e} \Omega^{-1} m^{-1}$ for Ir/CoFeB(2.3 nm) and $-(3.5 \pm 0.4) \times 10^4 \frac{\hbar}{2e} \Omega^{-1} m^{-1}$ for Ir/Co(2.3 nm), along with smaller bulk values $\sigma_{FL}^{bulk} = (1.8 \pm 0.3) \times 10^4 \frac{\hbar}{2e} \Omega^{-1} m^{-1}$ and $-(1.6 \pm 0.2) \times 10^4 \frac{\hbar}{2e} \Omega^{-1} m^{-1}$, respectively. The sign of $\sigma_{FL}^{bulk}$ is opposite for the Ir/CoFeB and Ir/Co samples, similar to the opposite signs of bulk FLT found previously for Pt/Py [32] and Pt/Co [25]. However, the interface FLT has the same sign (negative) in both Ir/CoFeB and Ir/Co. In the Ir($t_{Ir}$>1.2)/CoFeB/Ir($t_{Ir}$ =1.2) control samples we observe negligible $\sigma_{FLT}^{interface}$ and a small but non-zero $\sigma_{FLT}^{bulk}$ (~$(0.9 \pm 0.1) \times 10^3 \frac{\hbar}{2e} \Omega^{-1} m^{-1}$) (Fig. 3(i)). This is consistent with full cancelation of interface spin-orbit torque between the top and bottom Ir/CoFeB interfaces, and partial cancelation of the bulk spin currents for Ir layers with different thicknesses.

One must be careful in evaluating evidence of interfacial field-like SOTs to check that the results cannot be explained by spatial inhomogeneities in the resistivity as a function of thickness. Such inhomogeneities might give rise to a nonuniform current flow within the FM layer that produces an additional non-zero net Oersted torque. One should also be suspicious of other potential errors in subtracting off the Oersted field due to current flow in the non-magnetic layer. We can rule out these possibilities for our samples because of the large magnitude of the interfacial field-like torque we observe. For our Ir(1)/CoFeB(2.3) samples, even if all of the current flowing through the device were localized at the interface above the CoFeB layer, this would produce a maximum in-plane Oersted field of only -0.087 mT/Volt, which is less than half the magnitude of the effective field that we measure corresponding to the FLT. (See details of this calculation and plots of the measured FLTs before Oersted-field subtraction in Supplementary Material [29]). We have also verified that the current-induced SOTs are negligible in symmetrically-sandwiched structures containing only a single CoFeB layer Ti(1.5)/MgO(1.5)/CoFeB(2)/MgO(1.5)/Ti(1.5).

Previous measurements of interfacial FLTs larger than the maximum possible Oersted torque have been reported in Ti/NiFe/$Al_2O_3$ samples [22]. Substantial interfacial FLTs have also been reported in SiN/CoFeB/Ta/$TaO_x$ samples [21], but measurements on Ta/CoFeB/MgO and TaN/CoFeB/MgO samples show much smaller values [7,33]. Kim et al. [34] have suggested that DMI and interface-generated spin-orbit torques can share a common origin, which is intriguing given the strong DMI from Ir. However, for Pt [24] (which is similar to Ir in providing a strong DMI), any interfacial FLT is at least a factor of 10 weaker than we find for Ir.

**Switching application**

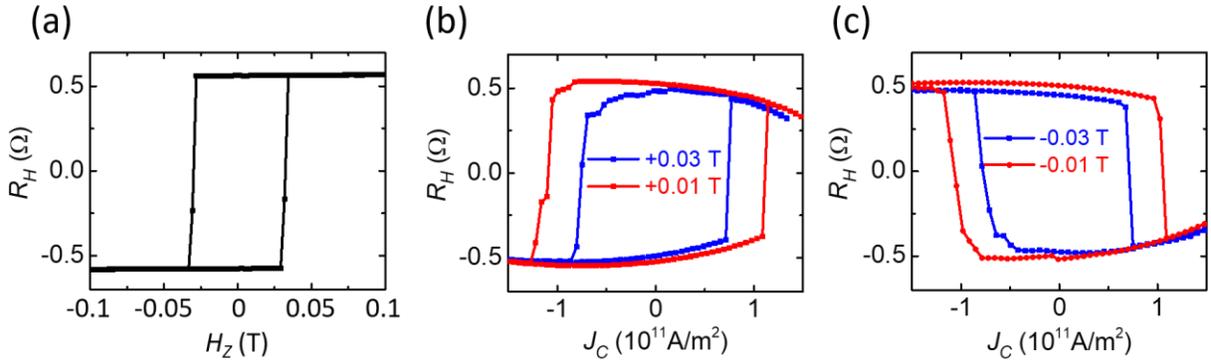

Fig. 4. (a) Anomalous Hall resistance of Ir(3)/CoFeB(0.8) as a function of an applied out-of-plane magnetic field. (b,c) SOT switching as a function of applied DC current density for two different orientations of a symmetry-breaking external magnetic field ($\mu_0 H_X = \pm 0.03$ Tesla and $\pm 0.01$ Tesla).

Finally we demonstrate current-induced SOT switching in a Ir(3)/Ti(0.4)/CoFeB(0.8)/Ti(0.2)/MgO(1.8)/Ti(1.5) heterostructure. The insertion of the 0.4 nm Ti dusting layer [35] between the Ir and the CoFeB enables perpendicular magnetic anisotropy (PMA) with a coercivity over 250 Oe (Fig. 4(a)) and an anisotropy field around 2.5 kOe. We observe switching of the PMA CoFeB in $20 \times 6$ μm$^2$ structures as the dc bias current is ramped (Fig. 4(b,c)) in the presence of a fixed symmetry-breaking in-plane magnetic field applied along current direction ($\mu_0 H_x = 0.01$-$0.03$ Tesla). The sign of the current-induced hysteresis loop is reversed upon reversing the direction of $H_x$, confirming that the mechanism is SOT switching. The minimum switching current density is just below $10^{11}$ A/m$^2$ (which corresponds to approximately 2 mA of applied DC current) at $\mu_0 H_x = 0.03$ Tesla, comparable to previous studies in other heavy metals [1–4,36].

**Summary**

We have measured both damping-like and field-like spin-orbit torque generated by sputtered Ir thin films. The damping-like torque is consistent with a conventional bulk spin Hall effect mechanism, with a a maximum spin torque conductivity $\sigma_{DL}^{eff} = (1.4 \pm 0.1) \times 10^5 \frac{\hbar}{2e} \Omega^{-1} m^{-1}$, a maximum spin torque ratio $\xi_{DL} = 0.042 \pm 0.005$, and an effective spin diffusion length approximately $\lambda_{Ir} = 1.6 \pm 0.3$ nm. Unlike most other heavy metals and heavy-metal alloys we have studied, the field-like torque has a dominant interface origin, with $\sigma_{FL}^{interface} = -(5.0 \pm 0.5) \times 10^4 \frac{\hbar}{2e} \Omega^{-1} m^{-1}$ for Ir/CoFeB(2.3 nm) and $-(3.5 \pm 0.4) \times 10^4 \frac{\hbar}{2e} \Omega^{-1} m^{-1}$ for Ir/Co(2.3 nm), each more than a factor of 2 stronger than the bulk contributions to the FLT. Although the SOTs generated by Ir are not as suitable for applications compared to Pt, Ta, and W, they are still sufficiently strong that they should be taken into account when incorporating Ir into SOT devices to achieve a strong Dzyaloshinskii-Moriya interaction.

**Acknowledgments**

We thank Ryan Tapping for the stimulating discussions. S.D is supported by Industrial Research and Consultancy Center (IRCC), IIT Bombay, Mumbai, India. A.B. is supported by the NSF through the Cornell Center for Materials Research (DMR-1719875). Sample fabrication and measurements were performed in part in the shared facilities of the Cornell Center for Materials Research and in the Cornell Nanoscale Science and Technology Facility, part of the National Nanotechnology Coordinated Infrastructure, which is supported by the NSF (NNCI-2025233).